\documentclass[runningheads]{llncs}
\usepackage[utf8]{inputenc} 
\usepackage[T1]{fontenc}    
\usepackage[hidelinks]{hyperref}       
\usepackage{url}            
\usepackage{booktabs}       
\usepackage{nicefrac}       
\usepackage{graphicx}
\usepackage{mathtools}
\usepackage{xcolor}
\usepackage{amsmath}
\usepackage{amssymb}
\usepackage{amsfonts}       
\usepackage{physics}
\usepackage{xspace}
\usepackage{bigints}
\usepackage{appendix}
\usepackage{cleveref}
\usepackage{mathtools}
\usepackage{subfig}
\usepackage{paralist}

\usepackage{tikz}
\usetikzlibrary{arrows.meta, positioning, shapes.misc}

\usepackage{algorithm}
\usepackage{algpseudocode}
\usepackage{braket} 

\newcommand{\R}{\ensuremath{\mathbb{R}}\xspace}

\newcommand{\hri}{Honda Research Institute Europe GmbH, Carl-Legien-Strasse 30, 63073 Offenbach am Main, Germany}
\newcommand{\ist}{Institute of Science Tokyo, 2-12-1 Ookayama, Meguro, Tokyo, 152-8530, Japan}
\newcommand{\aqa}{$\langle aQa ^L\rangle $ Applied Quantum Algorithms, Universiteit Leiden}
\newcommand{\liacs}{LIACS, Universiteit Leiden, Niels Bohrweg 1, 2333 CA Leiden, Netherlands}

\begin{document}
\title{Improving Quantum Multi-Objective Optimization with Archiving and Substitution}

\author{Linus Ekstr\o m\inst{1,4} \and
Takafumi Hosogi\inst{1,2} \and
Xavier Bonet-Monroig\inst{1} \and
Hao Wang\inst{3,4} \and
Thomas Bäck\inst{4} \and
Sebastian Schmitt\inst{1}
}
\institute{\hri \and 
\ist \and
\aqa \and
\liacs}

\authorrunning{L. Ekstr\o m et al.}
\titlerunning{Improving QMOO}
\maketitle

\begin{abstract}
Finding optimal solutions of conflicting objectives is a daily matter in many industrial applications, with multi-objective optimization trying to find the best solutions to them.
The advent of quantum computing has led to researchers wondering if the promised exponential advantage can be obtained for these problems by variational quantum multi-objective optimization (QMOO) algorithm.
Here, we improve it by introducing a Pareto Archiving and dominated solutions substitution, clearly improving in hyper-volume convergence at additional quantum and classical cost.
We propose the use of RMNK-landscapes as a unifying testbed for benchmarking QMOO, as it is common in classical multi-objective field.
By devising a generic classical-to-quantum mapping of these landscapes, we perform a numerical hyperparameter tuning of QMOO, significantly enhancing its performance.
Finally, we compare QMOO against well-known classical solvers for multi-objective tasks, NSGA-II/III, showing comparable results in small instances.
Our results demonstrate that QMOO, when carefully tuned for the task at hand, might be advantageous on harder problems than its classical counter-parts.

\keywords{Variational Quantum Multi Objective Optimization, Multi Objective Optimization, RMNK Landscapes, Algorithm Benchmarking}
\end{abstract}

\section{Introduction}\label{sec:introduction}
In recent years, progress in quantum computing has advanced steadily towards the point of practical utility. Although current hardware remains in the noisy intermediate-scale quantum (NISQ) era~\cite{Preskill2018quantumcomputingin}, improvements in qubit fidelity, coherence times, error correction and device scalability motivate research into potential useful use-cases of NISQ era machines. One area that has received considerable attention is combinatorial optimization. Specifically, hybrid variational algorithms such as the quantum approximate optimization algorithm (QAOA)~\cite{farhi2014quantum}. However, these algorithms were mostly focused on single objective optimization until recently~\cite{abbas2024challenges}. An interest in extending the variational quantum optimization field to problems with multiple competing cost functions can be found in the following works~\cite{chivilikhin2020mogvqemultiobjectivegeneticvariational,chiew2023scalarization,diezvalle2023multiobj-constraints,ekstromquantummultiobjective,kotil2025quantumapproximatemultiobjectiveoptimization}.

Problems involving multiple conflicting objectives arise naturally in virtually all domains of science, technology, and engineering.
From industrial design and logistics to finance, robotics, and resource scheduling, real-world applications rarely consider only a single objective. Decision makers must evaluate and navigate trade-offs between competing objectives, e.g.\ cost vs.\ performance or efficiency vs.\ robustness.
In such settings, optimizing for a single scalar objective is insufficient.
Rather, the goal requires the computation of the so-called Pareto-optimal solution sets, i.e.\ solutions for which no objective can be improved without degrading another~\cite{evolutionary-algorithms-coellocoello-book,multi-objective-deb-2001}.

In this work, we build upon the QMOO algorithm defined in~\cite{ekstromquantummultiobjective} by adding Pareto archiving and dominated substitution, resulting in a significantly improved algorithm for solving multi-objective optimization problems in terms of both quantum and classical computational resources.
We provide a first method for mapping any RMNK-landscape to a cost Hamiltonians implementable on quantum devices. We perform a hyperparameter tuning of the improved algorithm, and show that the algorithm works well with a low number of measurement shots and modest population size. Finally, we compare the tuned improved QMOO with NSGA-II/III on RMNK-landscapes of variable difficulty and find QMOO performs well with more complex optimization landscapes.

\section{Background}\label{sec:background}
\subsection{Multi-Objective Optimization}\label{sec:multiobjective}
Combinatorial multi-objective optimization is considers problems where a decision vector $\mathbf{x}\in\{0,1\}^N$ of $N$ binary variables must simultaneously minimize $M>1$ real-valued objectives
\begin{equation}
    \underset{\mathbf{x}\in\{0,1\}^N}{\text{minimize}}{}\hspace{5pt}\vec{C}(\mathbf{x}) = (C_1(\mathbf{x}), \cdots, C_M(\mathbf{x})).
\label{eq:minimization}
\end{equation}
The notion of an ordering from best to worst solutions can not be defined in this case because improving one objective may worsen another one. Instead, different solutions are compared through the
\begin{definition}[Pareto dominance]\label{def:paretodominance} A solution
$\mathbf{a}$ dominates $\mathbf{b}$, $\mathbf{a} \preceq \mathbf{b}$ iff.\ $a_i \leq b_i$ $\forall$  $i = 1, \dots, M$ and $\mathbf{a}\neq \mathbf{b}$.
The non-dominated solutions of the solution space are called the Pareto set, $\mathcal{PS}$, and their image under \eqref{eq:minimization} is called the Pareto front, which encodes all the best possible trade-off relations between the objectives.
\end{definition}

In practice, the goal of multi-objective optimization is to approximate the Pareto front.
The Pareto-optimal solutions are then passed to a decision-maker to choose the preferred ones based on various real-world scenarios and constraints which might not be fully captured by the objectives.

The general outline of multi-objective optimization algorithms begins with an initial set of solutions $X_{\text{initial}}\subset\{0,1\}^N$ and the corresponding objective vectors, $Y_{\text{initial}}\subset\mathbb{R}^K$ as an approximation of $\mathcal{PS}$.
The algorithm successively updates the guess, improving the approximation to the Pareto front (in objective space) for a specific update mechanism depending on the algorithm.

An important aspect of multi-objective optimization is the quality quantification of the sets of solutions against each other.
One such quality measure is the hypervolume (HV),
\begin{definition}[Hypervolume]\label{def:hypervolume}
    The hypervolume indicator of a point set $Y\subseteq \R^K$ is the volume of the region dominated by $Y$ and bounded above by a reference vector $\mathbf{r}\in \R^K$, i.e.\ $\operatorname{HV}(Y,\mathbf{r})=\lambda(\{\mathbf{y}\in\R^K \colon \mathbf{y} \prec \mathbf{r} \wedge \exists\mathbf{p}\in Y, \mathbf{p}\prec\mathbf{y}\})$, where $\lambda$ is the Lebesgue measure on $\R^K$. 
\end{definition}
The HV can be seen as the Lebesgue measure of the region dominated collectively by the approximation set.
Further, it is known to be \textit{Pareto compliant}~\cite{Falcon-CardonaM21,Falcon-CardonaE22,zitzler2003performance,ZitzlerBT06}, measuring both the convergence and coverage of the Pareto front.
For these reasons, it is the most important metric in multi-objective optimization, commonly used as to decide the set of solutions to keep.

\subsection{Quantum Multi-Objective Optimization}\label{sec:qmoo}
We adopt a hybrid quantum-classical variational algorithm for multi-objective optimization based on~\cite{ekstromquantummultiobjective}, where the authors have shown a general scheme which outputs a superposition state comprised of Pareto optimal solutions. The quantum circuit follows an alternating-layer ansatz composed of $L$ layers, each parametrized by angles \(\boldsymbol{\beta}^{l}\) and \(\boldsymbol{\gamma}^{l}\). The quantum state is evolved as
\begin{equation}\label{eq:layerunitary}
    \begin{split}
        \ket{\psi(\boldsymbol{\beta},\boldsymbol{\gamma})} & = U(\boldsymbol{\beta}^{L},\boldsymbol{\gamma}^{L})\cdots U(\boldsymbol{\beta}^{1},\boldsymbol{\gamma}^{1})\ket{\psi_0},\\
        U(\boldsymbol{\beta}^l,\boldsymbol{\gamma}^l) & =U_M(\beta^{l}_{K})\,P_{K}(\gamma^{l}_{K})\cdots U_M(\beta^{l}_{1})\,P_{1}(\gamma^{l}_{1}),\\     
    \end{split}
\end{equation}
where the initial state \(\ket{\psi_0}=\ket{+}^{\otimes N}\) is the uniform superposition of over all $2^N$ binary strings $x_i\in\{0,1\}^N$. Each layer $l$ consists of $K$ objective blocks, one for each objective function, where the $k$-th block applies a mixing unitary, $U_M(\beta_k)$, and a cost unitary, $P_k(\gamma_k)$.
\begin{equation}
    U_M(\beta_k)=\exp\Bigl[-i\beta_k\sum_{j=1}^{N} X_j\Bigr],
    \hspace{5pt}
    P_{k}(\gamma_k)=\exp\bigl[-i\gamma_k\,H_k\bigr],
\end{equation}
where $X_j$ is the Pauli-X operator on qubit $j$, and $H_k$ is the cost Hamiltonian encoding the $k$-th objective (see Sec.~\ref{sec:quantum-mapping}). The output state after $L$ layers takes the general form
\begin{equation}
\ket{\psi(\boldsymbol{\beta},\boldsymbol{\gamma})}=
    \sum_{\mathbf{x}\in\{0,1\}^{N}}
    \xi_{\mathbf{x}}(\boldsymbol{\beta},\boldsymbol{\gamma})\,
    \ket{\mathbf{x}},
\end{equation}
where the amplitudes $\xi_{\mathbf{x}}(\boldsymbol{\beta},\boldsymbol{\gamma})$ depend on the variational parameters. Sampling from this state produces a set of candidate solutions $\{\mathbf{x}_i\}$. 
At each iteration, the $N_{\text{most prob}}$ most probable solutions are extracted and evaluated under all objectives. The non-dominated subset forms a Pareto set approximation, whose quality is assessed via the hypervolume indicator (Sec.~\ref{sec:multiobjective}). The hypervolume is used as the cost function in a classical optimizer, which updates the variational parameters \(\boldsymbol{\beta}^{l}\) and \(\boldsymbol{\gamma}^{l}\) to improve the quantum circuit's sampling distribution. This defines a hybrid loop that iteratively refines the circuit towards concentrating probability mass on Pareto-optimal solutions by minimizing the negative hypervolume.

\subsection{RMNK Landscapes}\label{sec:RMNK}
RMNK-landscapes~\cite{Aguirre2004MNKlandscapes,aguirre2007MOEAonMNKlandscapes,verel2011objectivecorrelationMNK} generalize the classic NK-model~\cite{kauffman1989NKmodels} to the multi-objective setting. They represent an important class of scalable test problems with tunable complexity for combinatorial multi-objective optimization. 

\begin{figure*}
    \centering
    \raisebox{1.5mm}{\includegraphics[width=0.46\linewidth]{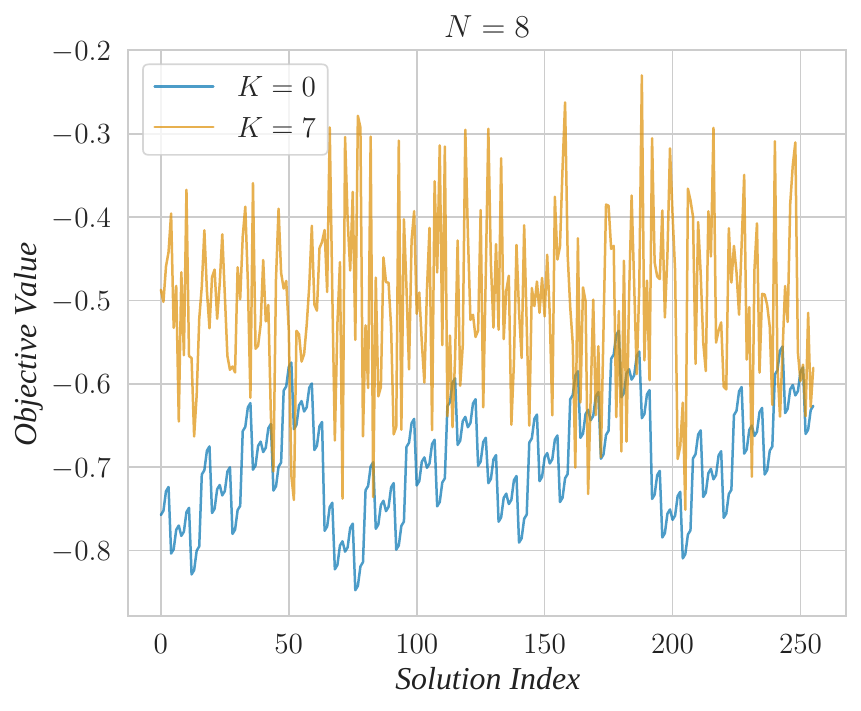}%
    }
    \hfill
    \includegraphics[width=0.49\linewidth]{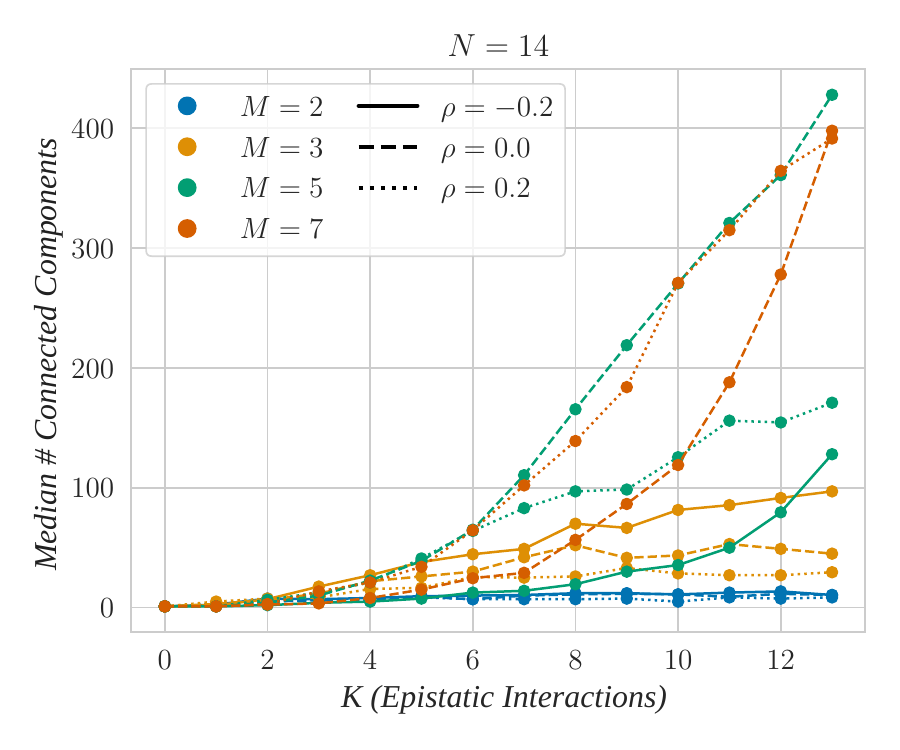}
    \caption{(Left) NK-landscape examples  for two different values of $K$. We see the same general structure, but increasing $K$ increases the ruggedness. (Right) Median number of bit-flip connected components in the Pareto front as a function of $K$ for ten different seeds per parameter setting. In general, the number of connected regions grows with $M$, $N$, and $K$ while varying with $\rho$.}
    \label{fig:connected_components}
\end{figure*}

The family of NK-landscapes is a problem-independent framework used to construct multi-modal landscapes. Originally proposed in the context of biology, the descriptors of the variables of the function family remain as such. $N$ refers to the number of (binary) genes in the genotype (i.e., the string length) and $K$ to the number of genes that influence a particular gene from the string (the epistatic interactions). 
The NK-landscapes have the property that, as the parameter $K\in\{0, N-1\}$ increases, the landscapes are gradually tuned from smooth to rugged as visible from an example shown in the left panel of as visible from an example shown in the left panel of Fig.~\ref{fig:connected_components}.

The fitness $f_{NK}(x)$ of a solution $x\in\{0, 1\}^N$ is calculated as  the mean value of $N$ fitness components $f_i$
\begin{equation}
\label{eq:nm_model}
    f_{NK}(x) = \frac{1}{N}\sum_{i=1}^Nf_i(x_i, x_{i_1}, \cdots, x_{i_K}).
\end{equation}

For each position $i$, its $K$ epistatic partners $i_1, i_2, \ldots, i_K$ are typically sampled u.a.r. from $[1...N]\setminus\{i\}$. The fitness function $f_i$ is defined by a lookup table $f_i\colon\{0,1\}^{K+1} \rightarrow \mathbb{R}$, which assigns a uniform random value from the interval $(0, 1)$ to every possible configuration of the bitstring made up from $x_i$ and its $K$ epistatic partners~\cite{kauffman1989NKmodels}.

In the multi-objective extension~\cite{Aguirre2004MNKlandscapes}, each objective function $f_{NK}^{(m)}$ is defined independently, with its own set of epistatic interactions and lookup tables,
\begin{equation}
    \forall m \in [1, M],\hspace{5pt} f_{NK}^{(m)}(x) = \frac{1}{N}\sum_{i=1}^Nf^{(m)}_{i}(x^{(m)}_i, x^{(m)}_{i_{1}}, \cdots, x^{(m)}_{i_{K}}).
    \label{eq:MNK-models}
\end{equation}
The overall landscape is vector valued, and solutions are compared using Pareto dominance as described in Sec.~\ref{sec:multiobjective}.

Another important step in the extension of the RMNK landscapes was introduced in ~\cite{verel2011objectivecorrelationMNK}, allowing explicit control of the pairwise correlation $\rho$ between objectives.  
For any two objectives $F_n$ and $F_p$ we have
\begin{equation}
    \mathrm{corr}(F_n,F_p)=
    \frac{1}{N^{2}\sigma_n\sigma_p}
    \sum_{i=1}^{N}\sum_{j=1}^{N}
    \operatorname{cov}(f_{ni},f_{pj}),
    \label{eq:rmnk_double_sum}
\end{equation}
where $f_{ni}$ is the contribution of locus $i$ to objective $n$ and $\sigma_n$ is the standard deviation of $F_n$.  
Because RMNK instances are generated so that contributions at different loci are independent, $\operatorname{cov}(f_{ni},f_{pj})=0$ for $i\neq j$. The double sum therefore collapses to,
\begin{equation}
    \mathrm{corr}(F_n,F_p)=
    \frac{1}{N^{2}\sigma_n\sigma_p}
    \sum_{i=1}^{N}
    \operatorname{cov}(f_{ni},f_{pi})
    =\rho .
    \label{eq:rmnk_single_sum}
\end{equation}
The parameter $\rho\in\bigl[-\tfrac{1}{M-1},1\bigr]$ thus tunes the desired degree of conflict ($\rho<0$) or alignment ($\rho>0$) between objectives.

We analyzed a wide range of RMNK landscapes in terms of the Pareto bit-flip connectedness Fig.~\ref{fig:connected_components}. 
By this we mean generating many different instances and checking whether the Pareto front is traversable by flipping only a single bit in the solution at a time. As $K$ increases, the Pareto-optimal solutions tend to become more widely scattered, which reduces the connectivity within the Pareto front and results in an increased number of disconnected components. Solutions belonging to different components cannot be reached from another using only single bit-flips, which presents a barrier for multi-objective algorithms.

\section{Mapping RMNK-Landscapes to Cost Hamiltonians}\label{sec:quantum-mapping}
We construct a quantum cost Hamiltonian from any given RMNK-landscape by mapping each local lookup table directly to a diagonal operator in the Pauli-$Z$ basis. The resulting operator reproduces the classical fitness values for all $2^N$ bitstrings. The cost Hamiltonian corresponding to the NK-landscape fitness function is expressed analog to Eq.~\eqref{eq:nm_model} 
as the sum of $N$ contributions 
\begin{align}
  \label{eq:cost-hamiltonian}
  H &=
  \frac{1}{N} \sum_{i=1}^{N} h_i \,,
\end{align}
where each component $h_i$ is supported on all subsets of $i$ and its epistatic partners $\{i_1, i_2,\ldots, i_K\}$:
\begin{align}
\label{eq:hi}
&h_i = \sum_{S\subseteq [1..K+1]} \alpha_i(S) Z_S, \quad Z_S=\bigotimes_{\ell \in S} Z_{p(\ell)} \\
&p(1) = i, \;p(2) = i_1, \;p(3) = i_2, \;\ldots,\; p(K+1) = i_{K}.
\end{align}
Note that, $h_i$ is a $(K+1)$-local Hamiltonian. We determine the coefficients $\alpha_i(S)$ such that the function values $f_i(x), x=(x_i, x_{i_1}, \cdots, x_{i_K})\in\{0,1\}^{K+1}$ are the eigenvalues of $h_i$, i.e., $\forall x\in \{0,1\}^{K+1}, \quad h_i\ket{x} = f_i(x)\ket{x}.$ To do so, we first consider the Hadamard transformation of the pseudo-Boolean function $f_i$:
\begin{equation}
f_i(x) = \sum_{S\subseteq [1..K+1]} \widehat{f}_i(S)\;\chi_{_S}(x),
\quad \chi_{_S}(x) = (-1)^{\sum_{j \in S}x_j}\,,
\end{equation}
where $\widehat{f}_i$ is the Hadamard transformation of $f_i$.
Note that the $Z_S$ operator naturally realizes the character $\chi_{_S}$ of the group $\{0,1\}^{K+1}$ (modulo 2 bitwise addition as group multiplication), i.e., $Z_S\ket{x}= \chi_{_S}(x)\ket{x}$. 
Hence, setting $\alpha_i(S) = \widehat{f}_i(S)$ will satisfy the eigenvalue problem $h_i\ket{x} = f_i(x)\ket{x}$,
\begin{equation}
    h_i\ket{x} = \sum_{S\subseteq [1..K+1]} \widehat{f}_i(S) Z_S\ket{x} = \sum_{S\subseteq [1..K+1]} \widehat{f}_i(S) \chi_{_S}(x)\ket{x} = f_i(x)\ket{x}.
\end{equation}
Finally, we have: 
\begin{equation}
\alpha_i(S) = \widehat{f}_i(S) = 2^{-(K+1)} \sum_{x\in\{0,1\}^{K+1}} f_i(x)\,\chi_{_S}(x).
\end{equation}

The construction above yields a diagonal Hamiltonian $h_i$ whose eigenvalues reproduce the local fitness table $f_i(x)$ for all bit-strings $x \in \{0,1\}^{K+1}$. Each coefficient $\alpha_i(S)$ in the Pauli-$Z$ expansion corresponds to a subset of interacting variables and is computed via a Walsh-Hadamard transform. Practically, we represent the function $f_i$ as a vector $\vec{f}_i = (f_i(b_1), f_i(b_2), \ldots, f_i(b_{2^{K+1}}))^\top$ with $b_i$'s are all computation basis. For instance, when $K=1$, we have $\vec{f}_i = (f_i(0,0), f_i(0, 1), f_i(1,0), f_i(1, 1))^\top$. The $\alpha_i$ coefficients can be computed as follows: 
\[
\alpha_i(S) = 
 \left[H_{K+1} \vec{f}_i\right]_r, \quad r = \sum_{j\in S}2^{j-1},
\]
where $[\cdot]_r$ takes the $r$-th row of a vector and $H_{K+1}  = H^{\otimes K+1}$ is the Hadamard matrix of size $2^{K+1}$. This yields a diagonal operator whose eigenvalues match the classical fitness values on all computational basis states.

Please note that the Hamiltonian of Eq.~\eqref{eq:cost-hamiltonian} and~\eqref{eq:hi} contains interaction terms up to order $K+1$ which cannot be implemented efficiently on most current quantum hardware where only two-qubit gates are realized. These higher order terms need to be decomposed into two-body terms (e.g.\ two-qubit gates) which involves large overheads~\cite{valianteLocalityReduction2021}. Only for $K=1$ the well-known quadratic Ising model is obtained which can be directly implemented with two-qubit gates.
But, for our investigation the implementability on current hardware is not so relevant since we focus algorithm performance, and employ exact state vector simulations where higher order terms can be readily implemented.

\section{Improving Quantum Multi-Objective Optimization Algorithm}
In this section we present the changes implemented in the improved QMOO algorithm: Pareto archiving and dominated substitution. 

\subsection{Adding a Pareto Archive}\label{sec:qmoo-archive}
A Pareto archive is a standard mechanism in multi-objective optimization that retains previously discovered non-dominated solutions to ensure progressive convergence. 
By preserving the best solutions found so far, the archive helps maintain or improve the quality of the Pareto approximation over the iterations. 
As shown in Algorithm \ref{alg:archive}, we integrate an unbounded archive into the QMOO framework, extending the original algorithm proposed  in~\cite{ekstromquantummultiobjective}  with only a small classical computation overhead.
This modification leads to a strictly improved variant. 
In our implementation, the hypervolume calculated based on the current archive population is passed to the classical optimizer. 

\begin{algorithm}[h]
\caption{QMOO with Pareto Archive. Here, $\mathcal{A}$ refers to the archive, $\mathcal{F}_{S}$ to the approximation to the Pareto front of the set $S$, and $T$ the maximal algorithm iterations.}
\label{alg:archive}
\begin{algorithmic}[1]
    \State Initialize $\mathcal{A} \gets \emptyset$, $\mathcal{F}_{\mathcal{A}} \gets [\,]$
    \For{$t = 1,\dots,T$}
        \State Sample QAOA state, obtain measurement counts, take top $N_{most\hspace{4pt}prob}$ bit-strings $\mathcal{S}_{\mathrm{cand}}$
        \State Discard any $s\in\mathcal{S}_{\mathrm{cand}}$ already in $\mathcal{A}$
        \If{$\mathcal{S}_{\mathrm{cand}}\neq\emptyset$}
            \State Compute $\mathcal{F}_{\mathrm{cand}} = \{\,f(s)\mid s\in\mathcal{S}_{\mathrm{cand}}\,\}$
        \Else
            \State $\mathcal{F}_{\mathrm{cand}} \gets [\,]$
        \EndIf
        \State $\mathcal{S}_{\mathrm{all}} \gets \mathcal{A} \cup \mathcal{S}_{\mathrm{cand}},\quad
               \mathcal{F}_{\mathrm{all}} \gets \mathcal{F}_{\mathcal{A}} \cup \mathcal{F}_{\mathrm{cand}}$
        \State \text{is\_pareto\_efficient}($\mathcal{F}_{\mathrm{all}})$
        \State Update $\mathcal{S}_\mathrm{all},\mathcal{F}_{\mathrm{all}}$ by selecting entries where $\text{mask}=\text{True}$
        \State Compute $HV = -\,\text{Hypervolume}(\mathcal{F}_{\mathrm{all}},\vec{r})$, return $HV$ to optimizer
    \EndFor
    \State \textbf{Output:} Final $\mathcal{S}_\mathrm{all}, \mathcal{F}_{\mathrm{all}}$
\end{algorithmic}
\end{algorithm}

\subsection{Dominated Solution Substitution}
A second improvement to the QMOO algorithm is called dominated solution substitution, which addresses a limitation in how candidate solutions are handled. In the original QMOO algorithm, the $N_{\text{most prob}}$ solutions are evaluated in terms of mutual non-domination, and solutions that are dominated are discarded. In the dominated solution substitution version of QMOO new candidate solutions are sequentially drawn from the measurement outcomes from the quantum circuit until we have a non-dominated set of size $N_{\text{most prob}}$. This ensures that the algorithm is always fed the same number of candidate solutions before the hypervolume calculation step. 

Substitution enhances early convergence by expanding the evaluated solution set. 
It may lead to a largely increased number of classical function evaluations as all sampled solutions might end up being evaluated if left unrestricted. 
While in general not being a problem, we avoid this from happening for the current scope of benchmarking.  
Therefore, we limit substitution candidates to at most $2 \times N_{\text{most prob}}$ bit-strings per iteration. This preserves the balance between number of bit-strings sampled and classical cost queries. 
\begin{algorithm}[h]
\caption{Dominated Solution Substitution}
\begin{algorithmic}[1]
    \State Initialize $\mathcal{S}_{\mathrm{sub}} \gets [\,]$, $\mathcal{F}_{\mathrm{sub}} \gets [\,]$
    \State Sample QAOA state, obtain measurement counts $\mathcal{S}_{\mathrm{raw}}$ sorted by probability
    \State $i \gets 1$
    \While{$|\mathcal{S}_{\mathrm{sub}}| < N_{\mathrm{most\hspace{4pt}prob}}$ \textbf{and} $i \leq \min\left(|\mathcal{S}_{\mathrm{raw}}|,\, 2 N_{\text{most prob}}\right)$}
        \State $s \gets \mathcal{S}_{\mathrm{raw}}[i]$
        \State Compute $f(s)$ and append $s$ to $\mathcal{S}_{\mathrm{sub}}$, $f(s)$ to $\mathcal{F}_{\mathrm{sub}}$
        \State Compute Pareto mask from $\mathcal{F}_{\mathrm{sub}}$
        \State Keep only non-dominated entries in $\mathcal{S}_{\mathrm{sub}}, \mathcal{F}_{\mathrm{sub}}$
        \State $i \gets i + 1$
    \EndWhile
    \State \textbf{Return} $\mathcal{S}_{\mathrm{sub}}$ as selected non-dominated candidates
\end{algorithmic}
\end{algorithm}

The substitution mechanism can be combined with the Pareto archive for further performance gains. In such a setting, dominated substitution is applied first to ensure the largest possible non-dominated candidate set. Then, these candidates are forwarded to the archive for potential inclusion. This ordering enables more efficient provision of bit-strings to the archive.   

\section{Results}\label{sec:results}
\subsection{Experimental Setup}\label{sec:experimental-setup}
\label{sec:setup}
To evaluate the performance of the improved QMOO, we conduct experiments on RMNK landscapes with $\rho = 0$, $M \in \{3, 5\}$, and $N \in \{12, 14, 24\}$. For $N = 14$ and below, we considered: $K \in \{1, N/2, N-1\}$ to investigate algorithm performance as a function of landscape complexity. For larger $N$, $K=1$ was used due to limitations in computational power. A hyperparameter sensitivity analysis was performed for the landscapes with $N=14$ (Fig.~\ref{fig:facetgrid-hyperparameter-search} and Fig.~\ref{fig:heatmap-fevals-to-99}). When $N_{\text{shots}} < N_{\text{most\ prob}}$, the value of $N_{\text{most\ prob}}$ has no practical effect, since QMOO can provide at most $N_{\text{shots}}$ solutions. In such cases, we simply used $1.0 \cdot N_{\text{pf}}$ as a representative setting, where $N_\text{pf}$ is the size of the true Pareto front. The number of layers used in the QMOO circuit was always set to $L=1$.

For comparison, several classical multi-objective optimization algorithms included in the Platypus~\cite{platypus} library were also tested. Specifically, we used NSGA-II and NSGA-III. Both NSGA-II and NSGA-III were extended to include a Pareto archive, similar to the one used in the quantum algorithm.

In order to fairly compare with QMOO using optimal values for $N_{\text{shots}}$ and $N_{\text{most prob}}$, optimal values were selected by performing a grid search over the following hyperparameters for NSGA-II/III: crossover probability $\in \{0.5, 0.7, 0.9, 1.0\}$, population size $\in \{20,\ 0.5 \cdot N_{\text{pf}},\ 1.0 \cdot N_{\text{pf}}\}$ (for NSGA-II), and division outer $\in \{4, 6, 8, 10, 12\}$ (for NSGA-III).

For the additional test cases with $N \in \{14, 24\}$, $K = 1$, and $M \in \{3, 5\}$, fixed hyperparameters were used. In these cases, QMOO was configured with $N_{\mathrm{\text{shots}}} = 1024$ and $N_{\mathrm{most\ prob}} = 20$. For NSGA-II and NSGA-III, the crossover probability was set to $1.0$, the population size for NSGA-II was $20$, and the division outer was set to $12$ for $M=3$ and $6$ for $M=5$ for NSGA-III. Ten runs were conducted for each individual setting.
The Powell optimizer from the \texttt{scipy}~\cite{2020SciPy-NMeth} library was used for optimizing the parameters of the quantum circuit.

\subsection{QMOO Improvements}
\begin{figure}
    \centering
    \includegraphics[width=0.49\linewidth]{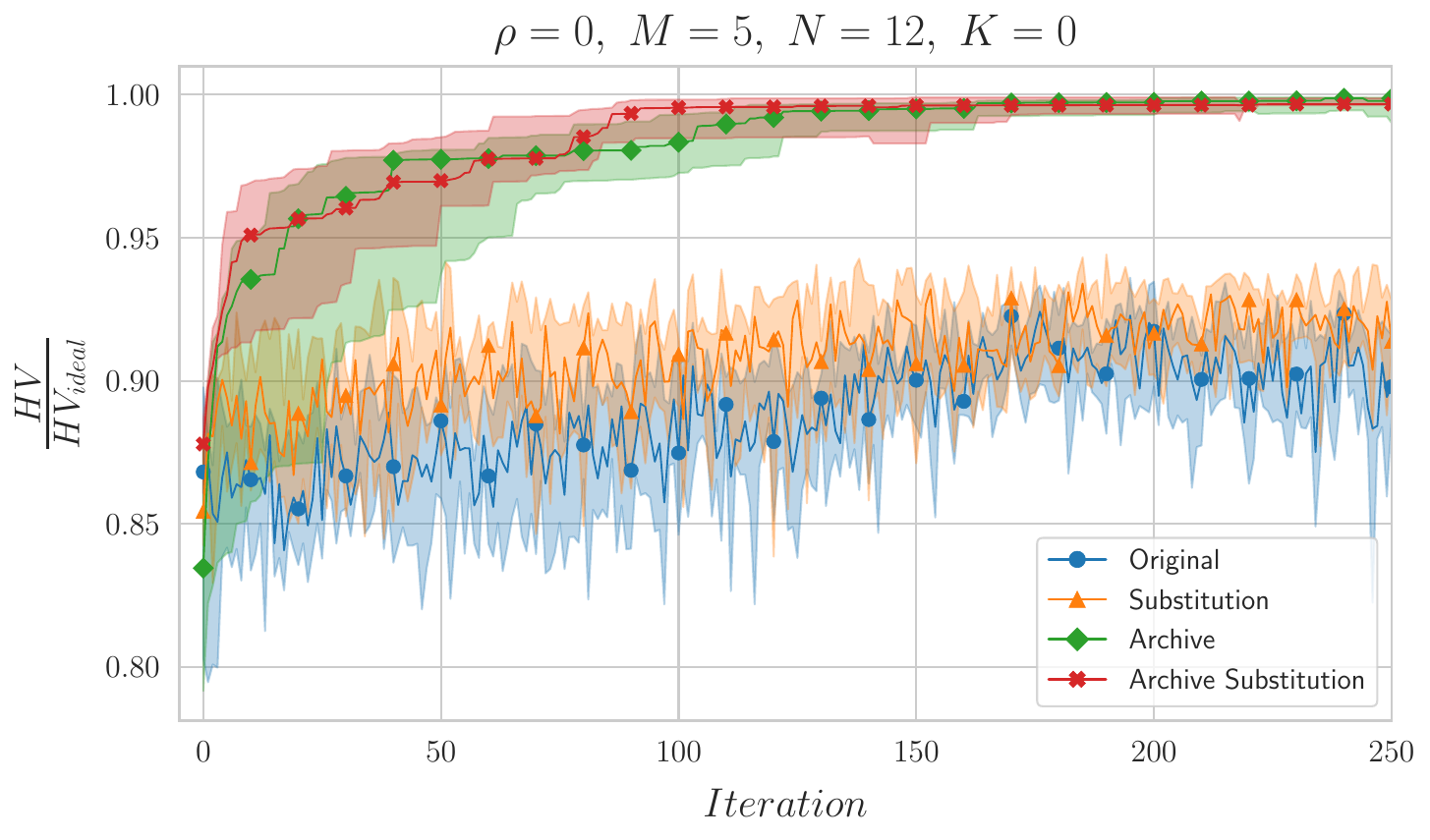}
    \includegraphics[width=0.49\linewidth]{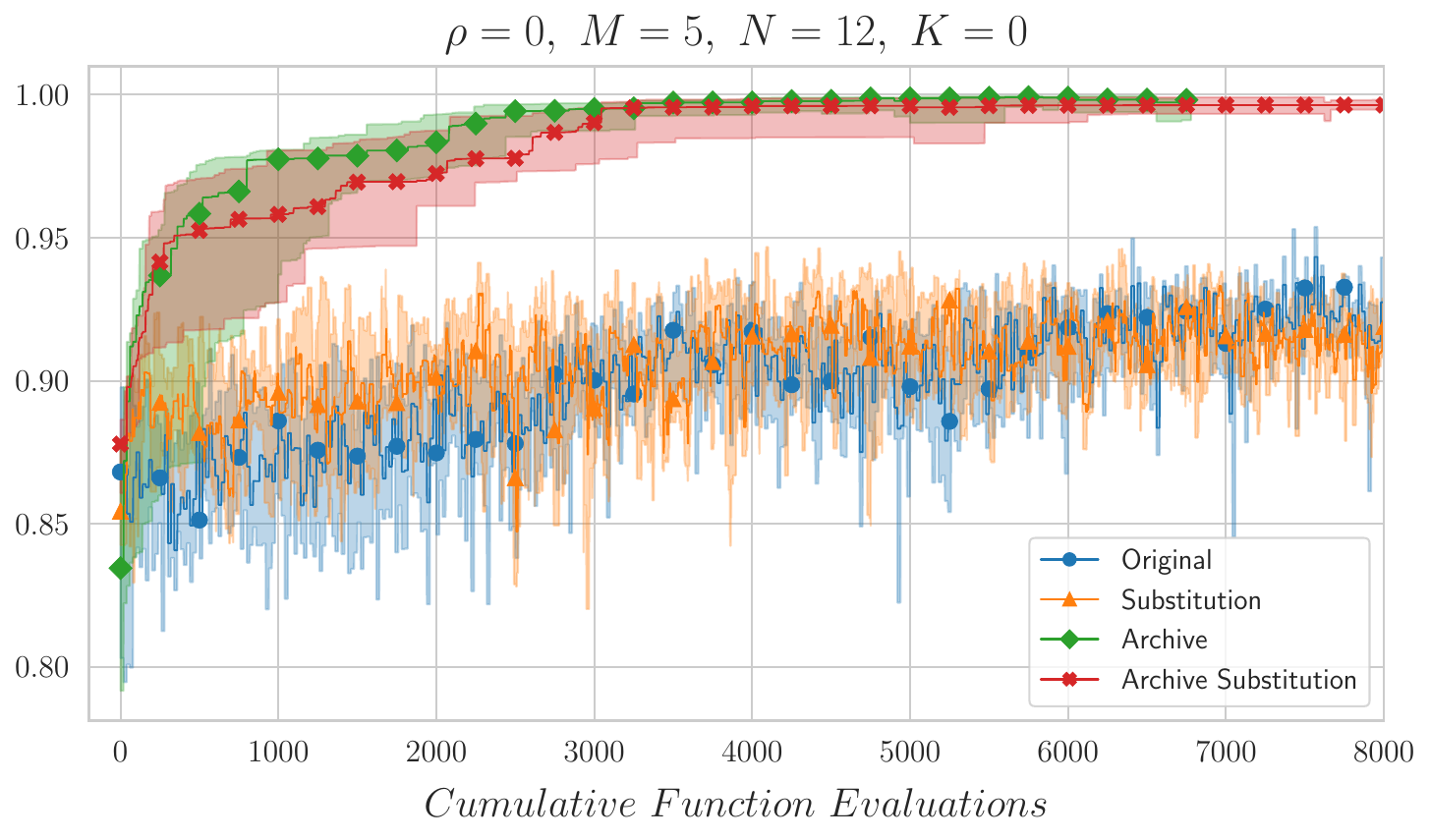}
    \caption{Improvement of normalized hypervolume HV/HV$_{ideal}$ of QMOO variants as function of  iterations (left) and  cost function evaluations (right) for one specific RMNK-landscape model with $M=5$ uncorrelated ($\rho=0$)  objectives, $N=12$ variables and $K=0$.}
    \label{fig:hypervolume-convergence-iter-fevals}
\end{figure}

First, we compare improvements for the convergence of the QMOO algorithm when considering archiving and substitution. As can be seen in Fig.~\ref{fig:hypervolume-convergence-iter-fevals} adding archiving improves the convergence ability of the algorithm significantly over the original QMOO. 
As for substitution alone, we observe a slight improvement especially in the early stages of the optimization, but overall its influence does seem to be small. Combining both archiving and substitution is slightly increasing the convergence in the early stage of the optimization  in terms of QMOO iterations, while no obvious trend is observed in terms of cost function evaluations.

\subsection{QMOO Hyperparameter Tuning}\label{sec:hyperparmeter tuning}
\begin{figure}
    \centering
        \includegraphics[width=1\linewidth]{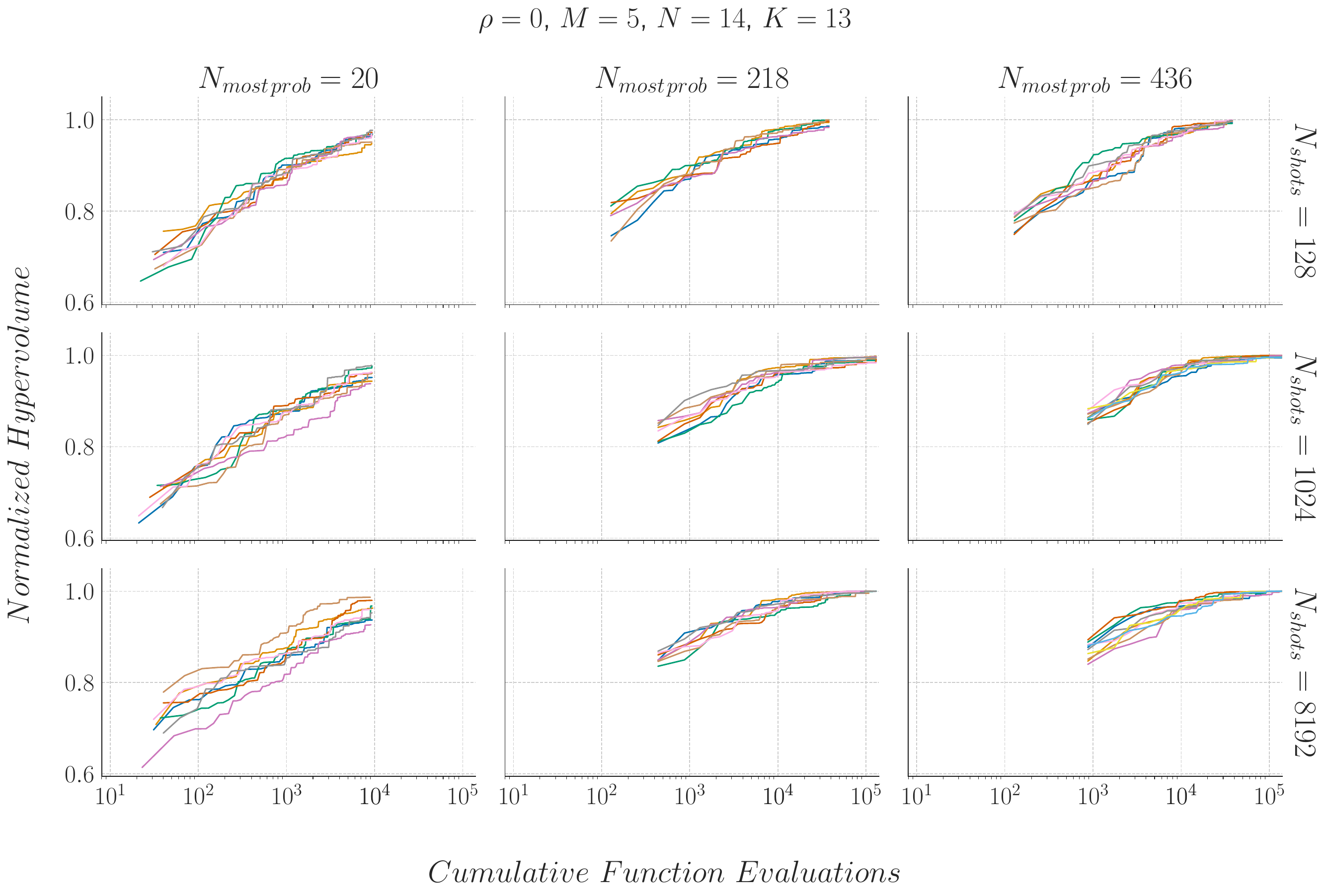}
    \caption{ 
    HV/HV$_{ideal}$ convergence of the QMOO with archiving and Pareto solution substitution for a selection of $N_{\text{most prob}}$ and $N_{\text{shots}}$ values. Note here the different starting points of the first iteration value as we observe the substitution can attempt to replace solutions up to $2\times N_{\text{most prob}}$ times resulting in a better average starting hypervolume. All cases use a fixed budget of $300$ QMOO iterations}
    \label{fig:facetgrid-hyperparameter-search}
\end{figure}

Using RMNK-landscapes, combinations of hyperparameters for the quantum optimization algorithm were tested. The hyperparameters investigated were the number of measurement shots $N_{\text{\text{shots}}}$ and the number of selected solutions $N_{most\hspace{4pt}prob}$, with $N_{\text{\text{shots}}} \in \{128, 1024, 8192\}$ and $N_{most\hspace{4pt}prob} \in \{20,\ 0.5 \cdot N_{\text{pf}},\ 1.0 \cdot N_{\text{pf}}\}$. Here, $N_{\text{pf}}$ denotes the size of the true Pareto front. 
Figure~\ref{fig:facetgrid-hyperparameter-search} shows the general convergence ability for QMOO with archive and substitution, as a function of function evaluations for various values of $N_{\text{shots}}$ and $N_{most\hspace{4pt}prob}$. 
As can be expected the number of function evaluations increases with more $N_{most\hspace{4pt}prob}$.
However,  the algorithm always obtains a good Pareto approximation and approaches the true Pareto front between 10$^4$ and 10$^5$ function evaluations  irrespective of hyperparameter values.

\begin{figure}
    \centering
        \includegraphics[width=1\linewidth]{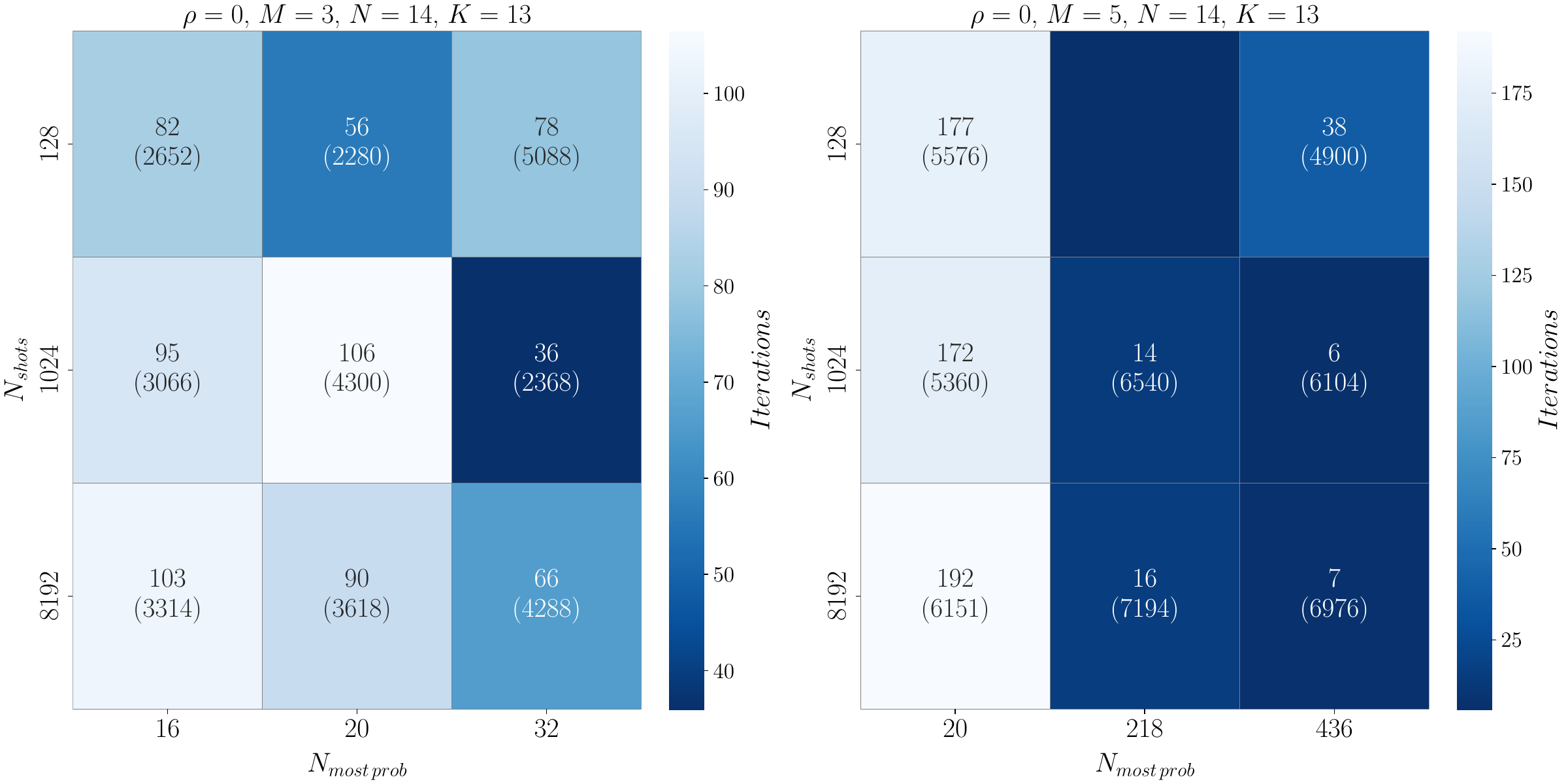}
    \caption{Hyperparameter tuning results for QMOO with archiving and Pareto solution substitution. For each problem setting, the numbers are the median number of iterations and function evaluations required to reach the threshold HV/HV$_\mathrm{ideal} \geq 0.95$. For empty cells no runs were performed as a larger $N_{\text{most prob}}$ setting supercedes it. 
}
    \label{fig:heatmap-fevals-to-99}
\end{figure}

Figure~\ref{fig:heatmap-fevals-to-99} presents the evaluation of each hyperparameter combination in terms of both iterations and function evaluations. The figure shows the number of iterations and function evaluations required to reach a threshold HV/HV$_{\textit{ideal}} = 0.95$.
We observe the effect of substitution increasing convergence speed in terms of iterations, while all values for $N_{most\hspace{4pt}prob}$ and $N_{\text{shots}}$ results in similar outcomes in terms of function evaluations. 

\subsection{Algorithm Benchmarking}
\begin{figure}
    \centering
        \includegraphics[width=1.\linewidth]{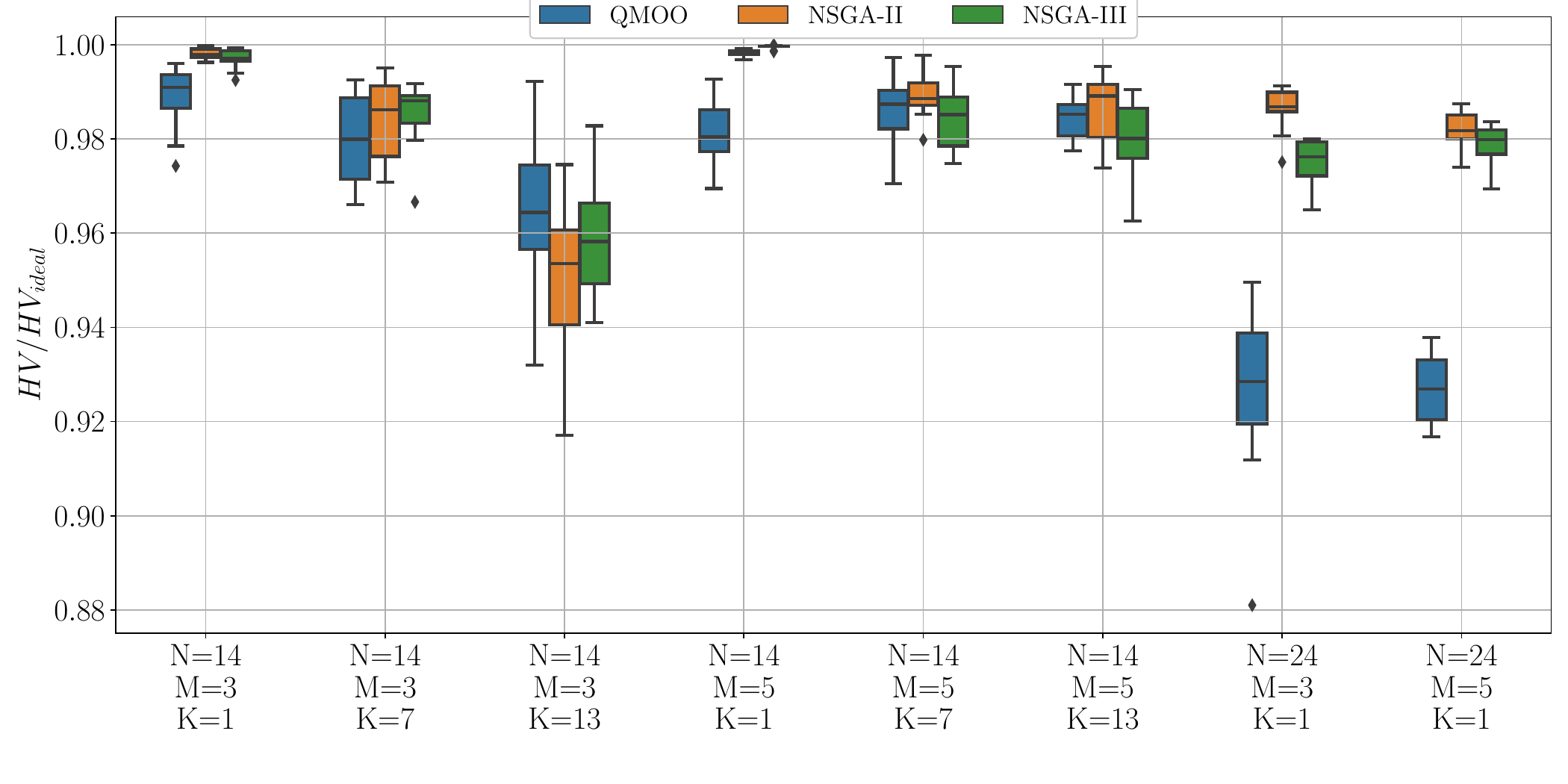}
        \caption{
        Normalized hypervolume values, using QMOO with archive and substitution, and NSGA-II/III with archive under different $(M, N, K)$. For the problem settings where hyperparameter tuning was conducted, the optimal set of hyperparameters was applied to each algorithm.
        }
    \label{fig:AUC}
\end{figure}

This section presents a comparative analysis between QMOO and NSGA-II/III. The results are shown in Fig.~\ref{fig:AUC}. For each problem setting, the maximum function evaluation is chosen as the fastest completed run among all algorithms. For the problem settings in which hyperparameter tuning was performed in Section~\ref{sec:hyperparmeter tuning}, the most effective combination of hyperparameters was selected based on heatmaps like in Fig.~\ref{fig:heatmap-fevals-to-99}. 

Specifically, among the hyperparameter combinations that achieved success rates above 50\%, the one with the smallest median number of function evaluations was chosen as the optimal setting. The target value of HV/HV$_{\textit{ideal}}$ was defined separately for each problem setting. Similar hyperparameter tuning procedures were also applied to NSGA-II and NSGA-III. For the remaining problem settings, fixed hyperparameters described in Section~\ref{sec:experimental-setup} were used.

In general we see comparable performance of QMOO across RMNK-landscape settings.
However, for larger problems we see a decline in the maximally achieved HV/HV$_{ideal}$ of QMOO relative to the  classical solvers NSGA-II/III, while the overall performance is still acceptable at above 90\%.  

\section{Discussion and Conclusion}\label{sec:discussion}

A central challenge in benchmarking quantum optimization algorithms like QMOO against classical baselines lies in defining exactly what is meant by \textit{fair} comparison. In classical algorithms, the cost function is queried when evaluating the candidate solutions within one generation. 
In contrast, running the quantum circuit generates many measurement shots where each measurement shot generates one potential solution candidate. However, only for a subset of these candidates the classical cost functions are evaluated. 
Therefore, we propose using the number of classical  function evaluations as an interpretable comparison metric. 

The addition of a Pareto archive constitutes the major algorithmic improvement proposed in this work. The QMOO performance is increased significantly for every instance we investigated (see Fig.~\ref{fig:hypervolume-convergence-iter-fevals} for an example).

Using the additional dominated substitution mechanism, we are able to trade off up convergence in terms of quantum circuit evaluations (i.e.\ algorithm iterations) with classical cost function evaluations by setting hyperparameters appropriately (see Fig.~\ref{fig:heatmap-fevals-to-99}). 
This is an important and valuable feature of the improved QMOO algorithm. It allows for tailoring performance according to available quantum and classical resources.
It is especially useful in current quantum hardware, where shifting cost to classical resources may significantly improve convergence ability. However, this becomes less important with cheaper quantum resources, or when the classical cost function becomes more costly to evaluate.

Our result suggest that hyperparameter tuning has a significant impact on the performance of QMOO. In this study, hyperparameter tuning was performed specifically for the cases where $K = N/2$ and $K = N-1$. Although similar hyperparameter tuning was conducted for both NSGA-II and NSGA-III, QMOO still achieved competitive or even superior performance in these tuned scenarios.

Although QMOO underperformed compared to NSGA-II/III for larger problems, it demonstrated superior performance in specific settings, particularly when the epistasis parameter $K$ was large (see Fig.~\ref{fig:AUC}). While large-scale problem instances with high $K$ have not yet been fully tested, the observed trends suggest that QMOO may outperform classical optimization algorithms in such scenarios. This hope is derived from the fact that the classical performance is degrading severely with increasing $K$ for large $N$ and $M$, while it appears that QMOO performance is not so much affected. 
However, at the current stage this is pure speculation and further investigations are required.

A natural next step is to study larger RMNK systems to verify whether the observed performance trends when increasing complexity hold as the problem size increases. 
Additionally,  investigating online adaptation policies for parameters $N_{\text{shots}}, N_{most\hspace{4pt}prob}$, and dominated substitution could be explored to dynamically balance quantum and classical resources during optimization, while  enhancing  QMOO algorithm performance  further. 
Finally, investigating impact of circuit-level noise and ensuring good performance under various noise models is an essential step  for bringing the QMOO algorithm to real quantum hardware.

In this paper, we have introduced two modification to the QMOO algorithm, namely Pareto archiving and dominated substitution, to improve its performance.
In order to benchmark such improvement, we provide a flexible mapping for arbitrary RMNK-landscapes to cost Hamiltonians, allowing one to easily modify the problems' hardness.
Taking advantage of this classical-to-quantum mapping, we show an improved performance of QMOO with increasing number of objectives, variables, and landscape complexities.
We further compare the improved QMOO algorithm against state-of-the-art classical algorithms for the same task, reaching comparable performance when the landscape complexity was high.
In summary, this manuscript constitutes the first step towards a systematic benchmarking of QMOO through RMNK-landscape, largely overlooked by the quantum community.
We anticipate the use of this technique to continue improving QMOO's algorithmic performance, and to overtake classical solvers in large-scale problems.

\begin{credits}
\subsubsection{\ackname} The authors acknowledge fruitful discussion with colleagues at the Honda Research Institute Europe, and extend their gratitude to A.~Bottarelli, V.~Dunkjo, P.~Hauke, M.~Olhofer, H.~Wersing, and B.~Sendhoff.
LE and SS acknowledge funding from the NeQST project of the European Union under Horizon European Program, Grant Agreement 101080086.
Views and opinions expressed are those of the author(s) only and do not necessarily reflect those of the European Commission. 
Neither the European Union nor the granting authority can be held responsible for them.
\end{credits}
\bibliographystyle{splncs04}

\end{document}